\documentclass[%
 aip,
 jmp,%
 amsmath,amssymb,
preprint,%
 reprint,%
groupedaddress]{revtex4-1}

\usepackage{graphicx}
\usepackage{dcolumn}

\usepackage{float}
\begin{document}


\title{Temporal Evolution of Social Innovation}
\author{Varsha S. Kulkarni}
 \affiliation{School of Informatics and Computing, Indiana University, Bloomington, IN 47408, United States}

\pdfoutput=1
\begin{abstract}
\small{Acceptance of an innovation can occur through mutliple exposures to individuals who have already accepted it. Presented here is a model to trace the evolution of an innovation in a social network with a preference $\lambda$, amidst topological constraints specified mainly by connectivity, $k$ and population size, $N_k$. With the interplay between properties of innovation and network structure, the model attempts to explain the variations in patterns of innovations across social networks. Time in which the propagation attains highest velocity depends on $\lambda^{-2}k^{-2}N_{k}^{1/2}$. Dynamics in random networks may lead or lag behind that in scale-free networks depending on the average connectivity. Hierarchical propagation is evident across connectivity classes within scale-free networks, as well as across random networks with distinct topological indices. For highly preferred innovations, the hierarchy observed within scale-free networks tends to be insignificant. The results have implications for administering innovations in finite size networks.}
%
\end{abstract}

\maketitle


Contagion is a mechanism often associated with the emergence of social innovations. An innovation connotes newness, be it in the form of a behavior, practice, product, or policy [1-3]. Statistical physics has found application in modeling innovations as social contagion processes [4-5]. As individuals in a social network collectively decide to accept or acquire an innovation, it leads to social reformations, trends or fashions. The success of technical, medical, policy, financial innovations is reflected in the trends observed [2,4,6,7,8]. Convergence to a particular phenomenon in social networks can occur through transmission of information from one individual to another [2,3,4,9,10]. While its modeling relies heavily on the principles of epidemic propagation, a susceptible individual often requires multiple interactions with acceptors of innovation, a form of social reinforcement, to accept the innovation [2-4]. Even after considering this aspect, variations in diffusion patterns of innovations across social networks continue to perplex researchers in various disciplines. Certain additional considerations may increase its credibility as a basis for modeling social contagion. Repeated interactions with acceptors typically increase individuals$^\prime$ awareness in stages [2-4], thereby persuading them to accept the innovation. As shown already, a very high number of intermediate stages can possibly delay the innovation acceptance indefinitely. In practice, however, this number remains arbitrary and depends on the social network. Moreover, interactions between individuals in social networks are restricted to occur with those they are connected to, who constitute their neighborhoods [10-14]. The connectivity pattern specifying the structure of a social network, therefore, is an important determinant of velocity of innovation propagation. A useful quantity studied is the emergence time $t_e$, defined as the time taken for attainment of maximum velocity in the network.\\
\indent The impact of network structure on propagation velocity is widely investigated [10-16]. An instantaneous rise in infected individuals is found in scale-free networks, wherein connectivity $k$ is governed by power law probability $P(k)\approx k^{-\gamma}$. The variance of connectivity diverges [15-16] when $2\leq\gamma\leq3$, which is also responsible for the unidirectional nature of propagation from higher to lower connectivity classes.  Additonally, random networks are connected relatively homogeneously. Epidemics can be curtailed therein if the average connectivity, $\langle k \rangle$ is low. However, little attention is focused on the way in which topological properties of a network ($k$, $P(k)$, network size) shape the hierarchical propagation dynamics of the social contagion process. Further, not every social network responds to an innovation to the same extent. Cultural, socio-economic attributes of a society contribute to its inertia that manifests in the unevenness of social change patterns across societies.  Apart from that, characteristics of the innovation (price or payoff) itself signify the society's preference for it, and are crucial for its acceptance [2,3,8,10,17]. It must be noted that experiments for studying performance of an innovation such as social reform policies or marketing of a product, are typically conducted on finite size social networks in a finite time horizon. Inclusion of all these features is thus imperative for a temporal analysis of social contagion.\\
\indent Here we address the complexity of the temporal evolution of innovation using a mathematical model that considers its interplay with the social environment. The model studies how an innovation emerges in response to the interplay of structural and social attributes of the network. It attempts to explain the variations in evolution across different societies in terms of the cumulative probability patterns. We compare the dynamics in three types of networks- scale-free network with $2\leq\gamma\leq3$ ($G_a$), $\gamma>3$ ($G_b$) and a randomly connected network ($G_r$). We verify that for an innovation preferred to the extent $\lambda$ in a network, the dynamics in $G_a$ dominates that in $G_b$ and $G_r$ provided $G_r$ is characterized by a sufficiently low $\langle k\rangle$. Irrespective of the network topology, $t_e$ for a class of connectivity $k$ shows power law variation with $k$, $\lambda$, and population of the class $N_k$. We find that the level of hierarchy is subject to the heterogeneity of connectivity. The resolution of this hierarchy has important implications on the selection of hubs for administering the propagation in finite size networks.\\
 \indent For studying the impact of the inherent peference for an innovation and network structure on the velocity of propagation, we consider a variant of the three-stage innovation acceptance model applied previously [4]. In this model, individuals can be in three discrete states, uninformed or susceptible (0), informed (1) and acceptor (2). $1$ is interpreted as a state of increased awareness of or willingness to accept an innovation. At any time $t$, for any connectivity $k$, the proportion of individuals in $0$, $1$, $2$ states is denoted by $p_{0}^{k}$, $p_{1}^{k}$, $p_{2}^{k}$ respectively. Here $p_{0}^{k}=N_{0}^{k}/N_k$, $p_1^k=N_1^k/N_k$ and $p_2^k=N_2^k/N_k$. $N_k$ is the number of individuals having connectivity $k$ and $N_0^k$, $N_1^k$, $N_2^k$ are numbers of those individuals in $0$,$1$,$2$ states respectively. 
Initially, the population consists of two types of individuals: an uninformed majority, and a very minute proportion of acceptors that take the lead to be different from the majority. Interaction of an uninformed individual with an acceptor can persuade the former to change irreversibly to informed state with a probability $\lambda$, equivalently, $0+2\overset{\lambda}{\rightarrow}1+2$. And similarly, $1+2\overset{\lambda}{\rightarrow}2+2$. Interactions between $0$ and $1$ do not produce any transitions. $\lambda$ is a parameter of the social network that specifies the propensity of any non-acceptor to change state upon interaction with an acceptor. It is a common attribute that represents the extent to which an innovation is preferred by individuals belonging to one society or group. Social networks often exhibit cohesiveness due to which individuals within a network tend to behave similarly. Hence, we consider the variation in $\lambda$ across individuals within a network as negligible when compared to that across different societies [2-3]. Information acquisition step delays the propagation. A lower value of $\lambda$ can add to the sluggishness of the process. $\lambda$ may also be interpreted as a socio-economic attribute that signifies the network$^\prime$s reservations about innovating. Tuning of the parameter on a continuous scale can result in a wider spectrum of time delays in propagation. With only an intermediate state (1), therefore, the model offers a general comparison between social contagion and the susceptible-infected model. Further, the interactions are constrained by the network structure as they only occur between individuals connected to each other. The rate equations are 

\begin{eqnarray}
\dot{p_{0}^{k}}=-\lambda kp_{0}^{k}\Theta,\nonumber\\
\dot{p_{1}^{k}}=\lambda kp_{0}^{k}\Theta-\lambda kp_{1}^{k}\Theta,\nonumber\\
\dot{p_{2}^{k}}=\lambda kp_{1}^{k}\Theta. \label{qg1}
\end{eqnarray}

$\Theta=\frac{\sum\limits_{k} (k-1) P(k)p_{2}^{k}}{\langle k \rangle}$ gives the average probability that a link from an uninformed individual points to an acceptor. 

We solve the equations for every class with the help of an auxillary function $\tau_k= \int_0^t p_2^k(t') dt'$, also referred to as internal time [4]. At the initial stages, the acceptors are very few in number, and we assume $\tau_k\approx \tau$. Thus $\int_{0}^{t} \Theta dt =\tau\mu$ where $\mu= \frac{\langle k \rangle-1}{\langle k \rangle}$ [15]. Using initial conditions $p_0^k(0)=1-\rho$, $p_1^k(0)=0$, and $p_2^k(0)=\rho$, we obtain the solutions as
\begin{eqnarray}
p_{0}^{k}=(1-\rho)e^{-\lambda\mu k\tau},\nonumber\\
p_{1}^{k}=(1-\rho)\lambda\mu k\tau e^{-\lambda\mu k\tau},\nonumber\\
p_{2}^{k}=1-(1-\rho)(\lambda\mu k\tau+1)e^{-\lambda\mu k\tau}.
\end{eqnarray} 

For any level of connectivity, the temporal increase in number of acceptors is accompanied by a decrease in the number of uniformed individuals. The number of informed individuals increases to a maximum and decreases thereafter. Patterns like these for all classes are similar to those shown in previous research [2-4]. The velocity of propagation is highest when the social network is maximally informed about the innovation. This happens for a class of connectivity $k$ when $dp_1^k/d\tau=0$, and the corresponding point of inflection is $\tau=1/\lambda\mu k$ . The proportion of acceptors at this point forms the critical mass [2-5] which in this case is 26$\%$ for every class. It suggests that the innovation emerges faster among individuals having relatively more connections than those having fewer connections. 
The time an innovation takes to propagate to the critical mass in any class or the network is defined as the emergence time of innovation for that class or the entire network. This is the time at which the innovation begins to propagate the fastest in the network. We compare propagation dynamics on different network structures on the basis of emergence time 

\begin{equation}
t_e^k=\int_{0}^{1/\lambda\mu k}\!\frac{dx}{p_2^k(x)}=\int_0^{1/\lambda\mu k}\!\frac{dx}{1-(1-\rho)(1+\lambda\mu kx)e^{-\lambda\mu kx}}.
\end{equation}

Substituting $x=y\sqrt\rho$, the integral in the limit $\rho<<1$ becomes

\begin{equation}
t_e^k=\frac{2}{\lambda^2\mu^2 k^2 \sqrt\rho}\int_0^{1/\lambda\mu k\sqrt\rho}\!\frac{dy}{(2\lambda^{-2}\mu^{-2}/k^2+y^2)}.
\end{equation}

A similar treatment has been applied previously [4]. Here, if $\rho=1/N_k$, then
\begin{equation}
t_e^k= 2\lambda^{-2}\mu^{-2} k^{-2}\sqrt{N_k} tan^{-1}\sqrt{N_k/2}.
\end{equation}
For a very large population, $\rho<<1$ the emergence time approximates to 
\begin{equation}
t_e^k=\pi \lambda^{-2}\mu^{-2} k^{-2}\sqrt{N_k}.
\end{equation}

The above approximation is valid for all classes as $N\rightarrow\infty$. In practice however, for finite size networks, it is restricted to a certain range of connectivity so that $N_k$ is sufficiently large. The parameter $\lambda$ determines the momentum of the process and affects all connectivity class alike. Eq.(6) implies that for all types of networks, $t_e^k$ varies as a power law in $k$, $N_k$ and $\lambda$. Square root dependence on $N_k$ is more apt in the infinite size limit (as $N\rightarrow\infty$). In practice, when network size is finite, density dependence of $t_e^k$ is through $\sqrt{N_k}tan^{-1}\sqrt{N_k/2}$, which is increasing in $N_k$. More specifically, the dynamics is divided for finite size networks as

\[
    t_e^k\approx 
\begin{cases}
    \pi\lambda^{-2}\mu^{-2}k^{-2}\sqrt{N_k} ,  &\text{if} (\lambda k) < (\lambda k)^{*};\\
                                              &                           N_k>N_k^{*}\\
    \pi\lambda^{-2}\mu^{-2}k^{-2}\sqrt{N_k}tan^{-1} \sqrt{N_k/2},              & \text{otherwise}
\end{cases}
\]
The values $(\lambda k)^{*}$, $N_k^{*}$ are associated with the (gradual) regime shift in the validity of the expressions. Evidently, the marginal effect of $N_k$ on $t_e^k$ (given by d$t_e^k$/d$N_k$) increases as $N_k$ decreases. Thus classes of higher connectivity perform much better when their densities are lower. This becomes crucial in case of finite size scale-free networks particularly with $2\leq g\leq 3$, wherein the presence of hubs facilitates faster propagation throughout the network. Further, a marginal increase in $k$ at lower levels impacts $t_e^k$ more when $\lambda$ is not high. This enhances the hierarchy at lower levels of connectivity. 

\begin{figure}[H]
\scalebox{0.35}{\includegraphics{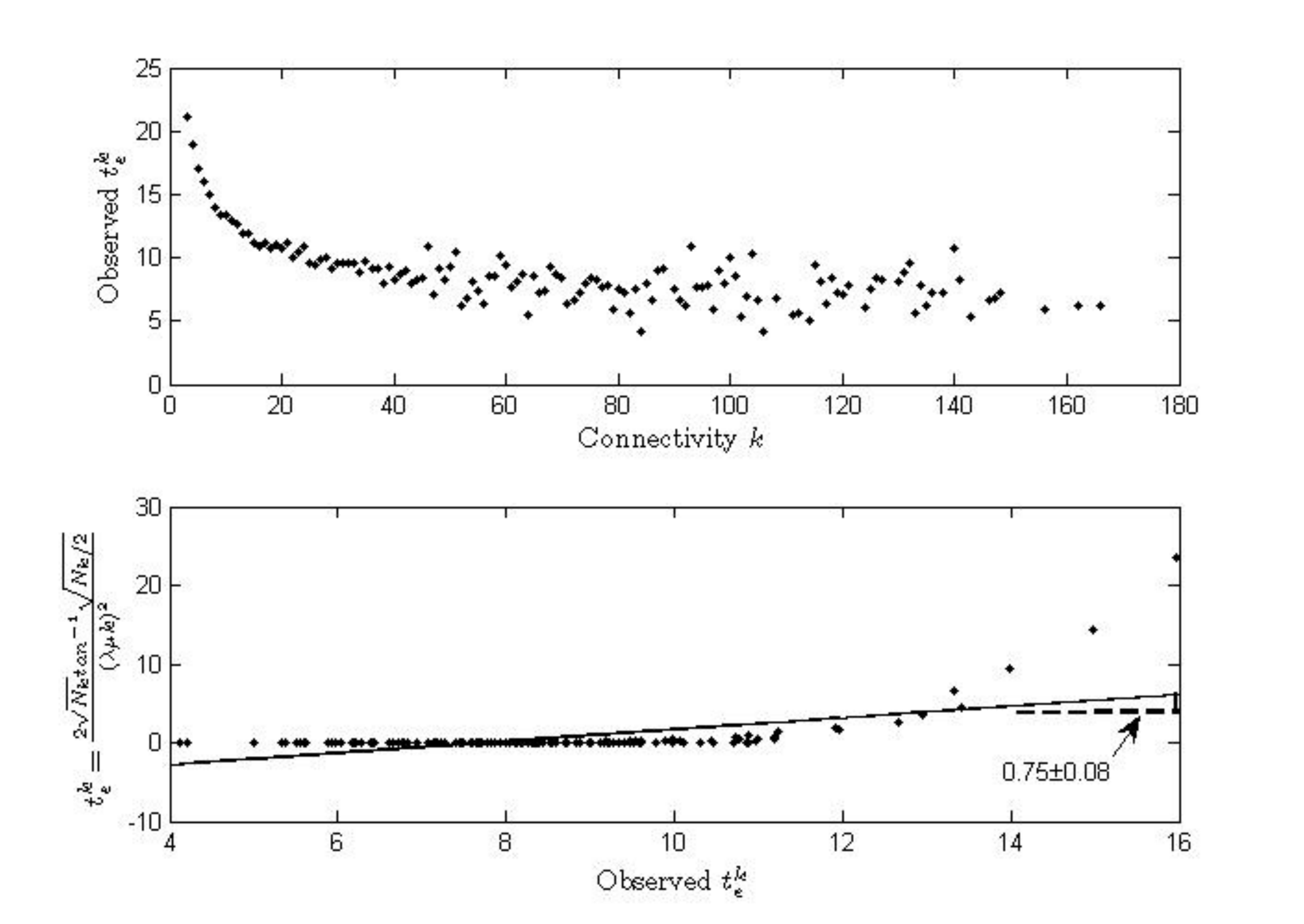}}
\caption{(a). Variation of $t_e^k$ with $k$ for the scale-free network $G_a$ with $\gamma\approx 2.35\pm 0.04$, $\langle k \rangle\approx10.5$ and $N=1.5\times 10^4$, for $\lambda=0.4$(Top). (b) Plot of observed with computed $t_e^k$ for the same network. Slope of straight line is $0.75\pm 0.08$ and the statistical fit to the observations is significant at the level 0.05 (Bottom).  }
\end{figure}

Heterogeneity is defined as $\langle k^2 \rangle/\langle k \rangle$, and it increases with $\langle k^2 \rangle$ or the variance of connections [18]. When $2\leq \gamma\leq 3$, the second moment of the power law distribution diverges in the limit $N\rightarrow\infty$. For finite size networks, however, the heterogeneity tends to be very high. In this context then, the result implies that heterogeneity of connections translates into a hierarchy in emergence times or velocities across different classes. We verify the relation by constructing a network of agents of type $G_a$, characterized by $P(k)\approx k^{-\gamma}$, $\gamma\approx 2.35$. The model is simulated for every vertex at each time step. The temporal evolution of number of acceptors can be traced by averaging the dynamics over 100 realizations. Fig.1 shows the agreement between observed and theoretical emergence times to a reasonably good approximation.  The fit can improve for larger $N$. We compare the propagation dynamics on two other network structures, (i) $G_b$, a scale-free network with $\gamma\approx 3.8$, and (iii) $G_r$, a randomly connected network. At first, in increasing order of heterogeneity we have $G_r<G_b<G_a$. $G_r$ is characterized by a Poisson probability distribution with an average connectivity $\langle k \rangle$, same as its variance. The dynamics of the entire network is centered about individuals with this level of connectivity as they are in abundance. For a random network of size $N$ and average connectivity $\langle k \rangle$, the average emergence time is $\langle t_e \rangle = \pi\sqrt{N}\lambda^{-2}\langle k \rangle^{-2}$. This retrieves the previous result for the dynamics in absence of topological constraints.  
\begin{figure}[H]
\scalebox{0.4}{\includegraphics{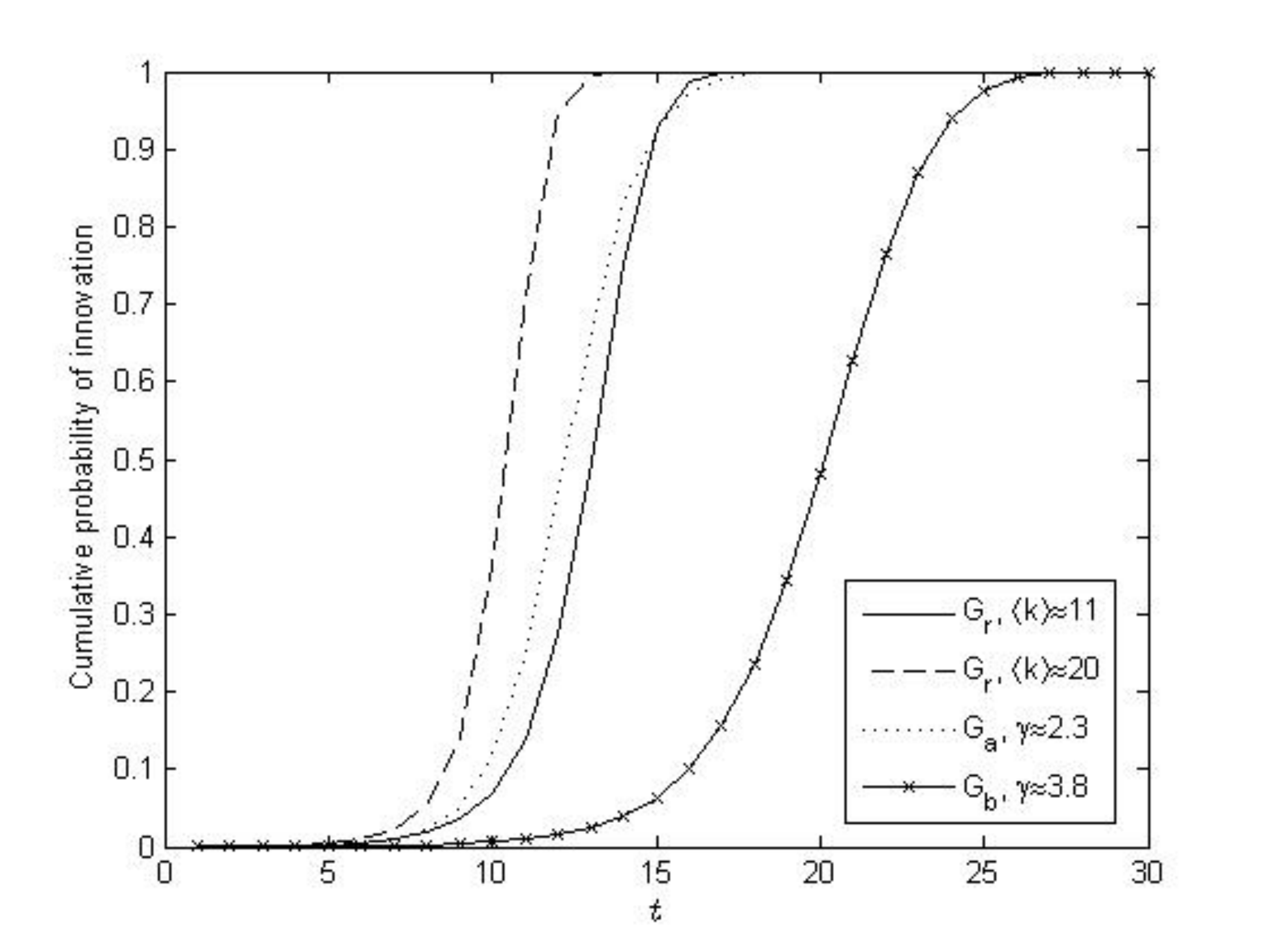}}
\caption{Comparison of four network confirgurations ($N=1.5\times10^4$)- $G_a$, $G_b$ ($\langle k \rangle\approx6$, $\gamma\approx3.8\pm0.04$), and $G_r$ with $\langle k \rangle\approx11, 20$ in terms of time variation of cumulative probability of innovation for $\lambda=0.4$. In the generation of random networks here, $k_{min}=0$, more than 1 acceptor may be required at the initial stages.}
\end{figure}
\begin{figure}[H]
\scalebox{0.4}{\includegraphics{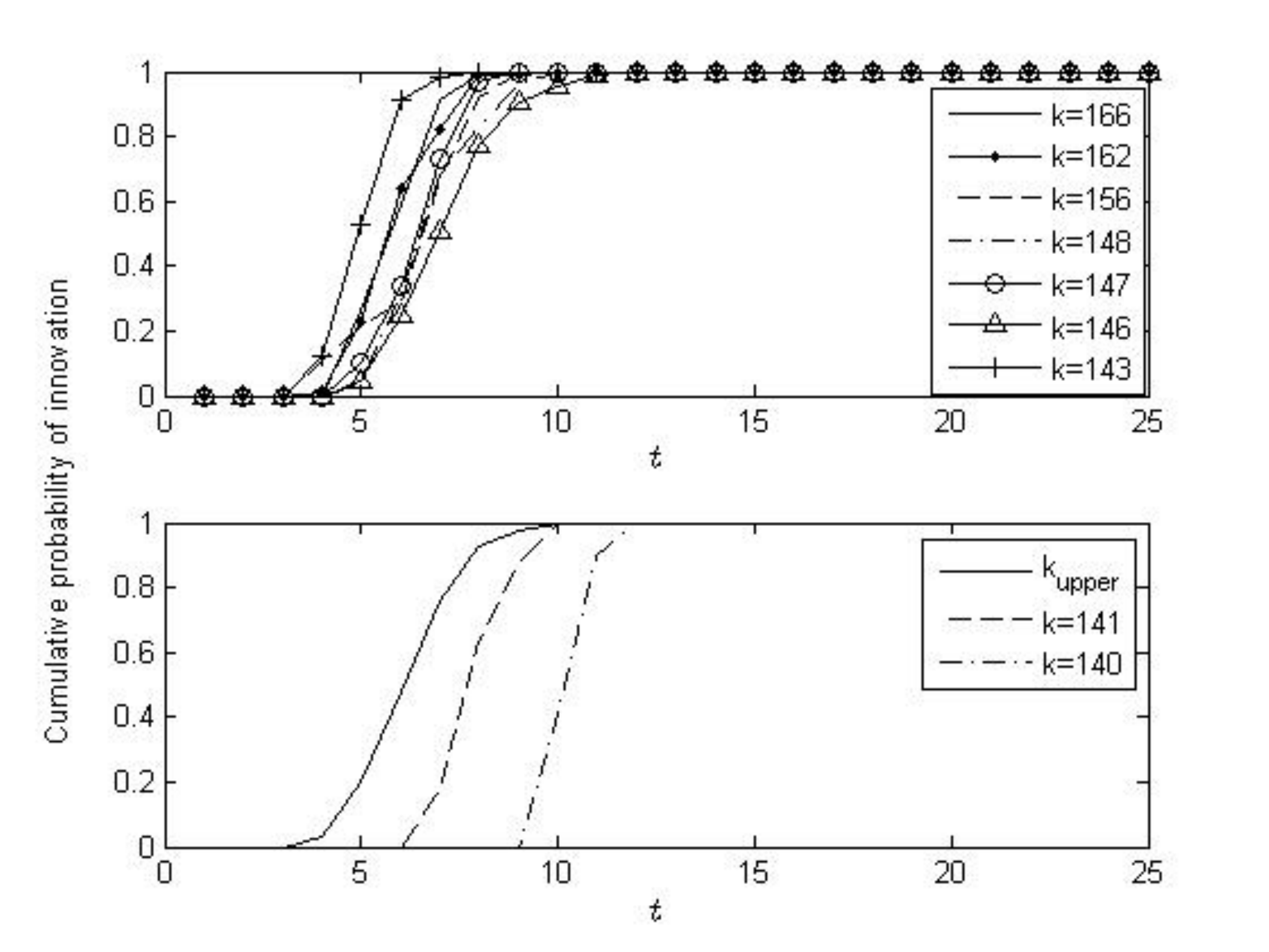}}
\caption{(a) Temporal evolution of cumulative probability of innovation for 7 classes of connectivity in $G_a$ (Top). (b) Average probability pattern of the 7 classes constituting the upper range $k_{upper}$ first order stochastically dominates all classes in the lower range. The dominance is significant as $D_c>D$ at 0.05 level (Bottom).}
\end{figure}
$G_b$ is characterized by $P(k)$ having well defined first and second moments, its heterogeneity of connectivity is much lower than in $G_a$. While maximum connectivity found in $G_b$ is lower, a large number of individuals is concentrated in classes of lower connectivity. This impedes the propagation significantly throughout the network. We compute the average emergence time for all networks as $\langle t_e\rangle= \sum_{k=k_{min}}^{\infty} t_e^k P(k)$. Assuming sufficiently large N, 
\begin{equation}
\langle t_e\rangle=\pi\lambda^{-2}\mu^{-2}\sum_{k=k_{min}}^{\infty} \sqrt{N_k}k^{-2}P(k)
\end{equation}
 For $G_a$, $G_b$, $P(k)=k^{-\gamma}/\zeta(\gamma, kmin)$, where $\zeta(\gamma, kmin)=\sum_{n=0}^{\infty}(n+k_{min})^{-\gamma}$ is the Riemann zeta function. And $G_r$ is governed by Poisson probability [18] $P(k)= e^{-\langle k \rangle}\langle k \rangle ^{k}/k!$.  This gives
\begin{equation}
\langle t_e\rangle=\pi\lambda^{-2}\mu^{-2}\sqrt{N} \frac{\sum_k k^{-3\gamma/2-2}}{(\zeta(\gamma, kmin))^{\frac{3}{2}}}
\end{equation}

for $G_a$ and $G_b$, and 
\begin{equation}
\langle t_e \rangle = \pi\lambda^{-2}\mu^{-2}\sqrt{N} e^{\frac{-3\langle k \rangle}{2}}\sum_k \frac{k^{-2}\langle k \rangle^{\frac{3k}{2}}}{k!^{\frac{3}{2}}}
\end{equation}

for $G_r$ with average connectivity $\langle k \rangle$. These expressions imply that for the same $\lambda$, $N$, propagation in $G_b$ will on an average lag behind that in $G_a$. The dynamics in $G_r$ is sensitive to changes in $\langle k \rangle$. Fig.2 shows that as $\langle k \rangle$ increases, $G_r$ makes the transition from lagging behind $G_a$ to leading it. Fig.2 confirms the correspondence of this behavior in $G_a$, $G_b$, $G_r$. Slight departures may be attributed to the finite size of networks, some approximations, and a few empirical deviations from the structural assumptions.

A question of interest is whether the difference between the response of $G_a$, $G_b$, and $G_r$ is significant? This is crucial in gauging the relative uncertainty associated with an innovation in a network. Moreover, hierarchical propagation appears more relevant for highly heterogeneous networks. In line with epidemiological literature, it is believed that only a few individuals with disproportionately high connectivity dominate the propagation throughout the network. For finite sized networks $G_a$ constructed here, while the overall behavior resembles Eq.(5), the marginal effect of increasing $k$ on $t_e^k$ not only diminishes but also fluctuates in the range of higher connectivity (Fig.1a). We quantitatively examine the difference made by presence of hubs in a network using stochastic dominance. 

Let the probability of an innovation (characterized by $\lambda$) for network structures $d$, $d^{\prime}$ be denoted by $p_d$, $p_{d^{\prime}}$ respectively. Then the dynamics in $d$ first order stochastically dominates that in $d^{\prime}$ if $p_d(t)\geq p_{d^{\prime}}(t)$ for all $t$ [18-19]. If the inequality holds for only some $t$, the dominance is said to be restricted. From Fig.3a, we find that the dominance of $G_r$ on $G_a$ and $G_b$ becomes strict as $\langle k \rangle$ increases.  Although, in this model, saturation is universal, of interest is the comparison of the functioning of networks in a finite time. Faster emergence makes the future of an innovation in a network more certain. For instance, trends, fads or bubbles in stock market [8] are known to affect temporary price predictability and investor behavior. Eq.(6) indicates the dynamics is deterministically hierarchical across connectivity classes.  So, $p_2^k(t)\geq p_2^{k^\prime}(t)$ for all $t$ when $k\geq k^{\prime}$. This means that the class of individuals with highest connectivity is supremely dominant. Further, the significance of the hierarchy can be inferred using Kolmogorv-Smirnov test statistic, $D(t)= \underset{t}{sup} |p_2^k-p_2^{k^\prime}|$, which is compared with the critical value $D_c$ at the level $\alpha$. We can verify in case of $G_a$ network of finite size, stochastic fluctuations hinder the observability of such a hierarchy especially in the range of higher connectivity. Fig.3b shows that it is the collective average dynamics of individuals in this range that is supremely and significantly dominant in the network, rather than any single individual or a class. In this way, the test specifies the range of connectivity for what can be considered as hubs in the network in that they dominate the dynamics. Fig.4 shows that the dominance is only restricted or absent if $\lambda$ is increased.

\begin{figure}[H]
\scalebox{0.4}{\includegraphics{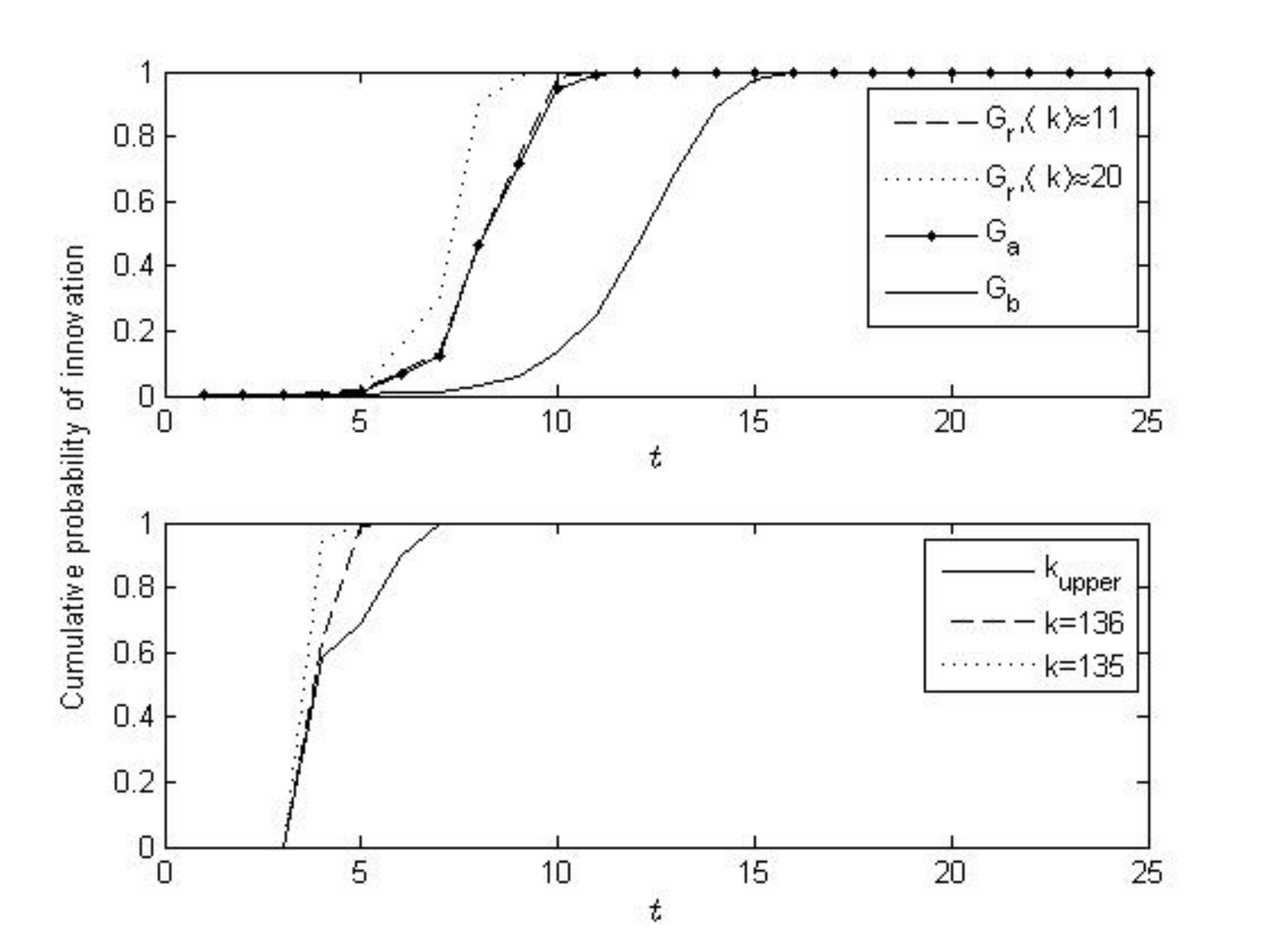}}
\caption{(a) Comparison of four network configurations $G_a$, $G_b$, and $G_r$ with $\langle k\rangle\approx11, 20$ in terms of time variation of cumulative probability of innovation for $\lambda=0.8$ (Top). (b) The average probability pattern for 10 classes constituting the upper range $k_{upper}$ does not singificantly dominate the lower range classes. The dominance remains insignificant for a broad upper range of connectivity (Bottom).}
\end{figure}

If only connectivity were important for administering the propagation in $G_a$, the precision of its range would be arbitrary. This is because the densities of the connectivity classes in the upper range are low enough to be sensitive to local effects. These effects may include immediate neighbors and their connections, clustering, and any fluctuations [21]. While the observed variance of connectivity in the tail of $P(k)\approx k^{-2.35}$ is very high, individuals possessing connectivity in this range tend to be as good as one another (or even behave anomalously). It is not surprising given that in Eq.(5), $tan^{-1}(\sqrt{N_k/2})$ diminishes the impact of a marginal increase in $k$ (on $t_e^k$) in this range. The present estimation and observation indicate that making a certain selection of few individuals with high connectivity ($N_k$, $k$), hinders the resolution of hierarchy that can be observed in this range. In contrast, to observe a particular desired level of hierarchy or significant dominance, the range of topological indices ($N_k$, $k$) required is wider, more uncertain. The meaning of ‘high’ and ‘few’ is relative and varies in a range that cannot be strictly defined while accounting for significant hierarchy. This is analogous to the quantum mechanical orbital of a subatomic particle whose position and velocity cannot be determined simultaneously with high precision [22]. A collective response of individuals in the higher range can be significantly dominant in the network, provided the innovation is not highly preferred (Fig.4a,b). Therefore, narrowing the range of topological coordinates to find a statistically significant hierarchical dominance in propagation may be misleading in finite size scale-free networks. \\
\indent Finally, the dominance of network structures in velocity of propagation is a relative concept. All structures are mere approximations and no particular one can absolutely facilitate or impede the propagation of innovation in a society. Connectivity in a social network signifies resources or access to information. If it is distributed unevenly, as in a scale-free network, the most resourceful individuals are among the first to accept an innovation. They may then ensure a faster propagation throughout all classes. Random networks, on the other hand, are more evenly distributed. There is competition for resources, which either facilitates or impedes the propagation for all individuals to a similar extent. With sufficient increase in resources on average, random networks would offer a more conducive environment for innovation. Network structure is crucial for administering the innovation.



\noindent\rule{4cm}{0.4pt}

\bibliography{References}
\renewcommand{\bibnumfmt}[1]{[#1] }

\end{document}